\documentclass[prc,preprint,showpacs,preprintnumbers,amsmath,amssymb]{revtex4}

\usepackage[mathscr]{eucal}
\usepackage{graphicx}
\def\slr#1{\setbox0=\hbox{$#1$}           
   \dimen0=\wd0                                 
   \setbox1=\hbox{/} \dimen1=\wd1               
   \ifdim\dimen0>\dimen1                        
      \rlap{\hbox to \dimen0{\hfil/\hfil}}      
      #1                                        
   \else                                        
      \rlap{\hbox to \dimen1{\hfil$#1$\hfil}}   
      /                                         
   \fi}

\def\be{\begin{eqnarray}}
\def\ee{\end{eqnarray}}

\renewcommand{\theequation}%
    {\arabic{section}.\arabic{equation}}
\makeatletter \@addtoreset{equation}{section} \makeatother

\begin{document}

\preprint{BCCNT: 04/111/328}

\title{Description of Gluon Propagation in the Presence of an $A^2$ Condensate}

\author{Xiangdong Li}
\affiliation{%
Department of Computer System Technology\\
New York City College of Technology of the City University of New
York\\
Brooklyn, New York 11201 }%

\author{C. M. Shakin}
\email[email address:]{casbc@cunyvm.cuny.edu}

\affiliation{%
Department of Physics and Center for Nuclear Theory\\
Brooklyn College of the City University of New York\\
Brooklyn, New York 11210
}%

\date{November, 2004}

\begin{abstract}
There is a good deal of current interest in the condensate
$<A_\mu^aA_a^\mu>$ which has been seen to play an important role
in calculations which make use of the operator product expansion.
That development has led to the publication of a large number of
papers which discuss how that condensate could play a role in a
gauge-invariant formulation. In the present work we consider gluon
propagation in the presence of such a condensate which we assume
to be present in the vacuum. We show that the gluon propagator has
no on-mass-shell pole and, therefore, a gluon cannot propagate
over extended distances. That is, the gluon is a nonpropagating
mode in the gluon condensate. In the present work we discuss the
properties of both the Euclidean-space and Minkowski-space gluon
propagator. In the case of the Euclidean-space propagator we can
make contact with the results of QCD lattice calculations of the
propagator in the Landau gauge. With an appropriate choice of
normalization constants, we present a unified representation of
the gluon propagator that describes both the Minkowski-space and
Euclidean-space dynamics in which the $<A_\mu^aA_a^\mu>$
condensate plays an important role.

\end{abstract}


\maketitle


\section{introduction}

Recently studies making use of the operator product expansion
(OPE) have provided evidence for the importance of the condensate
$<A_\mu^aA_a^\mu>$ [1-3]. (There is a suggestion that such a
condensate may be related to the presence of instantons in the
vacuum [4].) The importance of that condensate raises the question
of gauge invariance and there are now a large number of papers
that address that and related issues [5-19]. We will not attempt
to review that large body of literature, but will consider how the
presence of an $<A_\mu^aA_a^\mu>$ condensate modifies the gluon
propagator and the vacuum polarization function in QCD. We may
mention the work of Kondo [7] who was responsible for introducing
a BRST-invariant condensate of dimension two, \be
\mathcal{Q}=\frac{1}{\Omega}<\int
d^4x\,\mbox{Tr}\left(\frac{1}{2}A_\mu(x)A_\mu(x)-\alpha ic(x)\cdot
\bar{c}(x)\right)>, \ee where $c(x)$ and $\bar{c}(x)$ are
Faddeev-Popov ghosts, $\alpha$ is the gauge-fixing parameter and
$\Omega$ is the integration volume. Kondo points out that $\Omega$
reduces to $A_{min}^2$ in the Landau gauge, $\alpha=0$. Here the
minimum value of the integrated squared potential is $A^2_{min}$,
which has a definite physical meaning [7].

For recent discussion of the role of various vacuum condensates in
QCD one may refer to Refs. [20] and [21]. (In these works the
value given for the gluon condensate is \,\,\,\,\,\,
$<(\alpha_s/\pi) G^2>=0.009\pm 0.007$ GeV$^4$.) In our early work
[22] we assumed that the gluon condensate carried little or zero
momentum. The vector potential of the theory was divided into a
condensate field, $\mathbb{A}_\mu^a(x)$, and a fluctuating field,
$\mathcal{A}_\mu^a(x)$. The field $\mathbb{A}_\mu^a(x)$ is
independent of $x$ and has zero vacuum expectation value in our
model..

We define an order parameter, $\phi_0^2$, in a covariant gauge:
\be <vac\mid\mathbb{A}_\mu^a(0)\mathbb{A}_\nu^b(0)\mid vac> =
-\frac{\delta^{ab}}{8}\frac{g_{\mu\nu}}{4}\phi_0^2. \ee

The field tensor for QCD is given by \be
G^a_{\mu\nu}(x)=\partial_\mu A_\nu^a(x)-\partial_\nu A_\mu^a(x)
+gf^{abc}A_\mu^b(x)A_\nu^c(x).\ee We insert \be A_\mu^a(x) =
\mathbb{A}_\mu^a + \mathcal{A}_\mu^a(x) \ee into Eq. (1.3) and
define \be G^a_{\mu\nu}(x)=\mathbb{G}_{\mu\nu}^a +
\mathcal{G}_{\mu\nu}^a(x), \ee where \be \mathbb{G}_{\mu\nu}^a =
gf^{abc}\mathbb{A}_\mu^b\mathbb{A}_\nu^c \ee is the condensate
field tensor. We stress that, if the zero-momentum mode is
macroscopically occupied, $\mathbb{A}_a^\mu$ and
$\mathbb{G}_a^{\mu\nu}$ may be treated as classical fields.
However, we must maintain global color symmetry and Lorentz
invariance when using such fields.

As noted above, in our model [22], the gauge-invariant condensate
parameter\\
$<vac\mid:g^2(\tilde{\mu}^2)G_{\mu\nu}^a(0)G_a^{\mu\nu}(0):\mid
vac>$, is related to the condensate parameter,\\
$<vac\mid:g^2(\tilde{\mu}^2)A_{\mu}^a(0)A_a^{\mu}(0):\mid vac>$.
This relation follows from our assumption that the condensate is
in a zero-momentum mode. Since we have a phenomenological value
for
$<vac\mid:g^2(\tilde{\mu}^2)G_{\mu\nu}^a(0)G_a^{\mu\nu}(0):\mid
vac>$, obtained from QCD sum-rule studies [23], we can obtain a
value for
$<vac\mid:g^2(\tilde{\mu}^2)A_{\mu}^a(0)A_a^{\mu}(0):\mid vac>$ by
the following procedure.

Using the assumption that the condensate carries zero momentum, we
identify the condensate contribution as \be \nonumber
\frac{1}{4\pi^2}&<&vac\mid:g^2(\tilde{\mu}^2)G_{\mu\nu}^a(0)G_a^{\mu\nu}(0):\mid
vac> \\&=&
\frac{1}{4\pi^2}<vac\mid:g^2(\tilde{\mu}^2)\mathbb{G}_a^\mu(0)\mathbb{G}^a_\mu(0):\mid
vac> \\ &=&\frac{1}{4\pi^2}\sum f^{abc}f^{ab'c'}<vac\mid
g^4(\tilde{\mu}^2)\mathbb{A}_\mu^b(0)\mathbb{A}_\nu^c(0)\mathbb{A}_{b'}^\mu(0)\mathbb{A}_{c'}^\nu(0)\mid
vac>.\ee

We may write \be \nonumber &<&vac\mid
\mathbb{A}_\mu^b(0)\mathbb{A}_\nu^c(0)\mathbb{A}_{b'}^\rho(0)\mathbb{A}_{c'}^\eta(0)\mid
vac>\\
&=&\frac{\phi_0^4}{(32)(34)}[g_{\mu\nu}g^{\rho\eta}\delta_{bc}\delta_{b'c'}
+g_\mu^{\,\,\,\,\rho}
g_\nu^{\,\,\,\,\eta}\delta_{bb'}\delta_{cc'}+g_\mu^{\,\,\,\,\eta}
g_\nu^{\,\,\,\,\rho}\delta_{bc'}\delta_{cb'}].\ee

We have previously calculated matrix elements of this type by
several methods. In one work we calculated matrix elements of the
condensate potential after constructing $\mid vac>$ as a
coherent-state in the temporal gauge [24]. In another work [22] we
wrote  $\mathbb{A}_\mu^a(0)=\phi_0\eta_\mu^a$, where
$\eta_\mu^a\eta_a^\mu=-1$. In the latter scheme $\eta_\mu^a$ was
averaged over the gauge group when calculating matrix elements of
products of condensate fields. (One way to check the factor
(32)(34), which appears in the denominator of Eq. (1.9), is to set
$b=c, \mu=\nu, \rho=\eta$ and $b'=c'$ and sum over identical
indices.) We may insert the vacuum state between the operators to
obtain \be <vac\mid\mathbb{A}_\mu^b(0)\mathbb{A}_b^\mu(0)\mid
vac><vac\mid\mathbb{A}^\rho_{b'}(0)\mathbb{A}^{b'}_\rho(0)\mid
vac> =\phi_0^4.\ee This then agrees with the result obtained when
evaluating the right-hand side of Eq. (1.9).

Now, using Eq. (1.9) in Eq. (1.8), we find \be
\frac{1}{4\pi^2}<vac\mid:g^2(\tilde{\mu}^2)\mathbb{G}^a_{\mu\nu}(0)\mathbb{G}_a^{\mu\nu}(0):\mid
vac> =\frac{9}{(4)(34)\pi^2}(g^2(\tilde{\mu}^2)\phi_0^2)^2,\ee
from which we obtain \be
g^2(\tilde{\mu}^2)\phi_0^2=1.34(\mbox{GeV})^2.\ee (Here we use the
renormalization point $\tilde{\mu}^2\simeq 1$ GeV$^2$.) We will
make use of these results in the following.

In this work we discuss the form of the gluon propagator in some
detail. We also contrast the structure of the propagator in QCD
and QED. In this comparison the distinction between theories with
and without boson condensates is particularly clear. A
characteristic of a theory with condensates is the appearance of a
term proportional to $g_{\mu\nu}\delta^4(k)$ in the gluon
propagator. This term describes the macroscopic occupation of the
zero-momentum mode and provides a covariant representation of the
effect of the condensate in modifying the structure of the
propagator.

The organization of our work is as follows. In Section II we
review the introduction of the vacuum polarization tensor in the
case of QED. In Section III we discuss the vacuum polarization
tensor for QCD and in Section IV we review the Schwinger mechanism
for dynamical mass generation for gauge fields [25]. In Section V
we define a dielectric function for QCD and present the results of
our calculation of that quantity. In Section VI we provide values
of the gluon propagator in both Euclidean and Minkowski space and
make some comparison to the propagator obtained in lattice
simulations of QCD. Finally, Section VII contains some further
discussion and conclusions.

\section{The Photon Propagator and the Dielectric Function in
QED}

  In this section, we review standard results for the photon
  propagator in QED. (This material is available in the standard
  textbooks.) The propagator may be written as \be
  i\mathcal{D}^{em}_{\mu\nu}(k) = -i
  \left[\frac{(g_{\mu\nu}-k_\mu k_\nu/k^2)}{k^2(1-\Pi^1_{em}(k^2))}+
  \frac{\lambda k_\mu k_\nu}{(k^2+i\epsilon)^2}\right],\ee where
  $\Pi^1_{em}(k^2)$ is a finite quantity with the following limits
  (to order $\alpha$), \be \Pi^1_{em}(k^2) &=&
  -\frac{\alpha}{15\pi}\frac{k^2}{m^2},\qquad\qquad\qquad\qquad
   k^2\rightarrow 0 \,; \\ &=&
  \frac{\alpha}{3\pi}\ln\left(\frac{-k^2}{m^2}\right)-\frac{5\alpha}{9\pi},
  \qquad\qquad k^2\rightarrow -\infty. \ee It is useful to define the
  dielectric function, \be \kappa(k^2) = [1-\Pi^1_{em}(k^2)] \ee so
  that \be \kappa(k^2) \rightarrow
  \left[1+\frac{\alpha}{15\pi}\frac{k^2}{m^2}+\cdot\cdot\cdot\right]\,,\,\,\,\,\,
  k^2\rightarrow 0, \ee and \be \kappa(k^2)\rightarrow
  \left[1-\frac{\alpha}{3\pi}\ln\left(\frac{-k^2}{m^2}\right)+\frac{5\alpha}{9\pi}+\cdot\cdot\cdot\right],\,\,\,\,k^2\rightarrow
  -\infty. \ee

  A charge placed in the vacuum gives rise to a potential, \be
  V(\vec{k}) = \frac{e}{\kappa(\vec{k}^2)\,\mid \vec{k}\mid^2}, \ee
  which, for small distances, behaves as \be V(\vec{k})
  \rightarrow\frac{e}{[1-\frac{\alpha}{3\pi}\ln(\frac{\vec{k}^2}{m^2})+\cdot\cdot\cdot]
  \mid\vec{k}\mid^2}. \ee This is the standard result, which
  indicates that QED becomes strongly coupled at short distances.
  We make the observation that $D_{\alpha\beta}^{em}(k)$ has a pole
  at $k^2$=0. We make this apparently trivial observation, since
  we wish to demonstrate that in our model there is no
  corresponding pole in the gluon propagator in QCD - that is, the
  gluon becomes massive via the Schwinger mechanism [25].

  For completeness, we note that the polarization tensor has the
  form, \be \Pi_{\mu\nu}^{em}(k) &=&
  \left(g_{\mu\nu}-\frac{k_\mu k_\nu}{k^2}\right)\Pi^{em}(k), \\\nonumber
  &=&(g_{\mu\nu}k^2-k_\mu k_\nu) \Pi^{em}_1(k), \ee and we have \be
  i\mathcal{D}^{em}_{\mu\nu}(k)=i\mathcal{D}^{0}_{\mu\nu}(k)+
  i\mathcal{D}^{0}_{\mu\rho}(k)\,[i\Pi_{em}^{\rho\eta}(k)]\,i\mathcal{D}^{em}_{\eta\nu}(k),\ee
   where \be i\mathcal{D}^{0}_{\mu\nu}(k)=-i\left[\frac{(g_{\mu\nu}-k_\mu
  k_\nu/k^2)}{k^2+i\epsilon}+\frac{\lambda k_\mu
  k_\nu}{(k^2+i\epsilon)^2}\right]. \ee

  It is also useful to rewrite Eq. (2.10) as \be
  i\mathcal{D}_{\mu\nu}(k)=i\mathcal{D}_{\mu\nu}^{(0)}(k)+i\mathcal{D}_{\mu\rho}^{(0)}(k)
  \left[\frac{i\Pi(k)}{1-\Pi(k)/k^2}\right]^{\rho\eta}i\mathcal{D}_{\eta\nu}^{(0)}(k).\ee
  The term $[-i\lambda k_\mu k_\nu/(k^2+i\epsilon)^2]$ is common
  to both sides. Therefore we may put \be \mathcal{D}_{\mu\nu}^T(k)
  \equiv -[g_{\mu\nu}-k_\mu k_\nu /k^2] \mathcal{D}_T(k), \ee
  \be \mathcal{D}_{\mu\nu}^{(0)T}(k) \equiv -[g_{\mu\nu}-k_\mu
  k_\nu/k^2]\mathcal{D}_T^{(0)}(k), \ee with \be \mathcal{D}_T(k)
  = \frac{1}{k^2-\Pi(k)+i\epsilon}, \ee and \be
  \mathcal{D}_T^{(0)}(k) = \frac{1}{k^2+i\epsilon}, \ee to obtain
  the relation \be \left[\frac{\Pi(k^2)}{1-\Pi(k^2)/k^2}\right] =
  [\mathcal{D}_T^{(0)}(k)]^{-1}\,[\mathcal{D}_T(k)-\mathcal{D}_T^{(0)}(k)]\,[\mathcal{D}_T^{(0)}(k)]^{-1}.\ee

  The left-hand side of Eq. (2.17) is related to the time-ordered
  product of the currents, \be \left(g^{\mu\nu}-\frac{k^\mu
  k^\nu}{k^2}\right)\,\left[\frac{\Pi(k^2)}{1-\Pi(k^2)/k^2}\right] = i\int
  d^4x\, e^{ik\cdot x}<vac\mid T[j^\mu(x)j^\nu(0)] \mid vac>.
  \ee  Note that $\Pi(k^2)$ is the irreducible self-energy, while the
  matrix element of the time-ordered product of the currents gives
  rise to a reducible form.

  We remark that the equation for the vector potential is \be
  \left[-g^{\mu\nu}\Box
  +(1-\frac{1}{\lambda})\partial^\mu\partial^\nu\right]\,A_\nu(x)=-j^\mu(x).
  \ee Thus \be \partial_\mu j^\mu(x) = \frac{1}{\lambda}\Box
  (\partial_\nu A^\nu(x)).\ee

  However, in QED we have \be \partial_\mu j^\mu(x) = 0 \ee as an
  operator relation from which it follows that \be \Box
  (\partial_\nu A^\nu(x)) =0 \ee is an operator relation.

  Since the current is explicitly conserved in QED, $\Pi(k^2)$ is
  independent of the gauge-fixing parameter. In general, we can write \be j^\mu(x) = j^\mu_T(x) +j^\mu_L(x),
  \ee where \be j^\mu_T(x) = (g^{\mu\nu}-\partial^\mu
  \frac{1}{\Box}\partial^\nu)j_\nu(x), \ee and \be j^\mu_L(x) =
  \partial^\mu\frac{1}{\Box}\partial^\nu j_\nu(x). \ee We see from Eq. (2.21), that $j_L^\mu(x)=0$ in QED.

  In the case of QCD the situation is more complicated since current
  conservation does not appear at the operator level in a
  covariant gauge. Gauge fixing breaks the general local gauge
  invariance of the theory. (However, gauge fixing is necessary
  for covariant quantization, since without such a procedure one
  finds that the momentum conjugate to $A_0^a(x)$ is zero.) In QCD
  one usually uses path-integral quantization. That formalism
  leads to the introduction of ghost fields. These fields insure
  unitarity for the gluon channels.

  Here we will follow Lavelle and Schaden [26] and define the
  nonperturbative gluon propagator \be
  \mathcal{D}^{non\,pert}_{\mu\nu}(k^2)=\mathcal{D}_{\mu\nu}(k^2)-\mathcal{D}^{pert}_{\mu\nu}(k^2).\ee
The nonperturbative propagator is transverse in any covariant
gauge. That is a consequence of the Slavnov-Taylor identities
which state that the full and the perturbative longitudinal
propagators are the same in any covariant gauge. Therefore, the
difference appearing in Eq. (2.26) yields a purely transverse
result. In the following discussion we will consider the
calculation of the nonperturbative gluon propagator and insure
that our propagator is transverse.

  \section{Covariant Quantization in QCD}

  Consider the Yang-Mills Lagrangian for a $SU$(3) color theory
  without quarks. We have \be \nonumber \mathcal{L}(x)
  =-\frac{1}{4}G_{\mu\nu}^a(x)G^{\mu\nu}_a(x)-\frac{1}{2\lambda}(\partial_\mu A^\mu_a(x))^2
  -\partial_\mu\bar{\phi}_a(x)\partial^\mu\phi_a(x)\\+ gf_{abc}[\partial_\mu\bar{\phi}_a(x)\phi_b(x)
  A_c^\mu(x)], \ee where $\phi_a(x)$ and $\bar{\phi}_a(x)$ are
  ghost fields and $\lambda$ is a gauge-fixing parameter. Now with
  \be G_a^{\mu\nu}(x)=\partial^\mu A_a^\nu(x)-\partial^\nu A_a^\mu(x)
  +gf^{abc}A_b^\mu(x)A^\nu_c(x), \ee and using Eq. (3.1), we have
  \be
  \partial_\mu G^{\mu\nu}_a(x)+\frac{1}{\lambda}\partial^\nu(\partial_\mu
  A_a^\mu(x))&=&j^\nu(x)\\\nonumber &=&
  -gf^{abc}A_\mu^b(x)G^{\mu\nu}_c(x)-gf^{abc}[\partial^\nu\bar{\phi}_b(x)]\phi_c(x),\ee
  which we write as \be \left[-g^{\mu\nu}\Box
  +(1-\frac{1}{\lambda}\partial^\mu\partial^\nu\right]\,A_\nu^a(x)=-J_a^\mu(x),\ee
  where \be
  J_a^\mu(x)=-gf^{abc}A_\nu^b(x)G_c^{\nu\mu}(x)-gf^{abc}\partial_\nu(A_b^\nu(x)A_c^\mu(x))
  -gf^{abc}[\partial^\mu{\bar{\phi}_b(x)}]\phi_c(x).\ee Note that
  $J_a^\mu(x)$ is conserved in the classical theory. We write
  \be J_a^\mu(x)=J_{T,a}^\mu(x)+J_{L,a}^\mu(x)\ee using definitions
  analogous to those in Eqs. (2.24) and (2.25). Then \be
  \partial_\mu J_a^\mu(x)=\partial_\mu J_{L,a}^\mu(x) \ee and \be
  \partial_\mu J_a^\mu(x) =
  \frac{1}{\lambda}\Box(\partial_\mu A^\mu_a(x)). \ee As in QED,
  the constraints imposed by current conservation in the physical
  Hilbert space can be maintained by calculating with
  $J_{T,\mu}^a(x)$ instead of $J_\mu^a(x)$. (We remark that the
  ghost fields do not contribute to $J_{T,\mu}^a(x)$.) Note that
  \be \Box A_{T,a}^\mu(x) = J^\mu_{T,a}(x), \ee \be
  \frac{1}{\lambda}\partial^\mu(\partial_\nu
  A_{L,a}^\nu(x))=J_{L,a}^\mu(x), \ee and \be
  \frac{1}{\lambda}\Box(\partial_\nu A^\nu_{L,a}(x))=\partial_\mu
  J^\mu_{L,a}(x), \ee which also follows from Eqs. (3.7) and
  (3.8).

  We may also write \be
  [g^{\mu\nu}-k^\mu k^\nu/k^2]\left[\frac{\Pi(k^2)}{1-\Pi(k^2)/k^2}\right]\delta_{ab}
  &=&[g^{\mu\nu}k^2-k^\mu
  k^\nu]\left[\frac{\Pi_1(k^2)}{1-\Pi_1(k^2)}\right]\delta_{ab}\\\nonumber
  &=& i\int d^4x\,e^{ik\cdot x}<vac\mid
  T[J^\mu_{T,a}(x)J^\nu_{T,b}(0)]\mid vac>, \ee where the
  polarization tensor is defined as \be
  \Pi^{\mu\nu}_{ab}(k)=(g^{\mu\nu}-k^\mu
  k^\nu/k^2)\Pi(k^2)\delta_{ab}, \\ =(g^{\mu\nu}k^2-k^\mu
  k^\nu)\Pi_1(k^2)\delta_{ab}.\ee

  The division of the field into transverse and longitudinal parts
  does not have the same utility in QCD as in QED, since
  $J_{T,a}^\mu(x)$ and $J_{L,a}^\mu(x)$ have a nonlinear
  dependence on the gluon field. Therefore, the field equations do
  not separate into transverse and longitudinal equations in the
  case of QCD. The quantities $\Pi(k^2)$ and $\Pi_1(k^2)$ are defined in
  terms of conserved currents, $J_{T,a}^\mu(x)$. If we work with
  $J_a^\mu(x)$ rather than $J_{T,a}^\mu(x)$, the ghost fields will
  insure that $\Pi^{\mu\nu}_{ab}(k)$ has a transverse structure.
  However, if one does not insure the constraint
  $<\psi\mid\partial_\mu A_a^\mu(x)\mid\psi>$ = 0, one finds a
  dependence on the parameter $\lambda$ in $\Pi(k^2)$. While the
  presence of ghosts insure unitarity relations, they do not serve
  to impose the constraint $\partial_\mu A_a^\mu(x) =0$. Our calculation
  corresponds to a diagrammatic analysis, made in
  the Landau gauge, with condensate ghosts added to insure the
  transverse nature of $\Pi^{\mu\nu}_{ab}(k)$ [26].

\section{dynamic mass generation via the schwinger mechanism}

The term in $J_a^\nu(x)$,
$-g^2f^{abc}f^{cde}A_\mu^b(x)A_d^\mu(x)A_e^\nu(x)$, will give rise
to a gluon mass term if there is a gluon condensate in the QCD
ground state. We find a contribution to the conserved (transverse)
current of the form \be
J_a^\mu(x)=-m_G^2\left[g^{\mu\nu}-\partial^\mu\frac{1}{\Box}\partial^\nu\right]\,\mathcal{A}_\nu^a(x),\ee
where \be m_G^2=\frac{9}{32}g^2(\tilde{\mu}^2)\phi_0^2.\ee This
corresponds to a contribution to the polarization tensor of the
form \be \Pi_{\mu\nu}^{ab}(k)=\delta_{ab}(g_{\mu\nu}-k_\mu
k_\nu/k^2)m_G^2,\ee with $m_G^2=614$ MeV, if we use Eq. (1.12).
[See Fig. 1]. It is, therefore, useful to define
\be\Pi_1(k^2)=\frac{m_G^2}{k^2}+\frac{\Pi_A(k^2)}{k^2},\ee where
the second term does not have a pole as $k^2\rightarrow 0$. We
also have \be \Pi(k^2)=m_G^2+\Pi_A(k^2).\ee The appearance of a
pole at $k^2=0$ in $\Pi_1(k^2)$ defines the Schwinger mechanism
[25].

\begin{figure}
\includegraphics[bb=40 25 350 500, angle=0, scale=1]{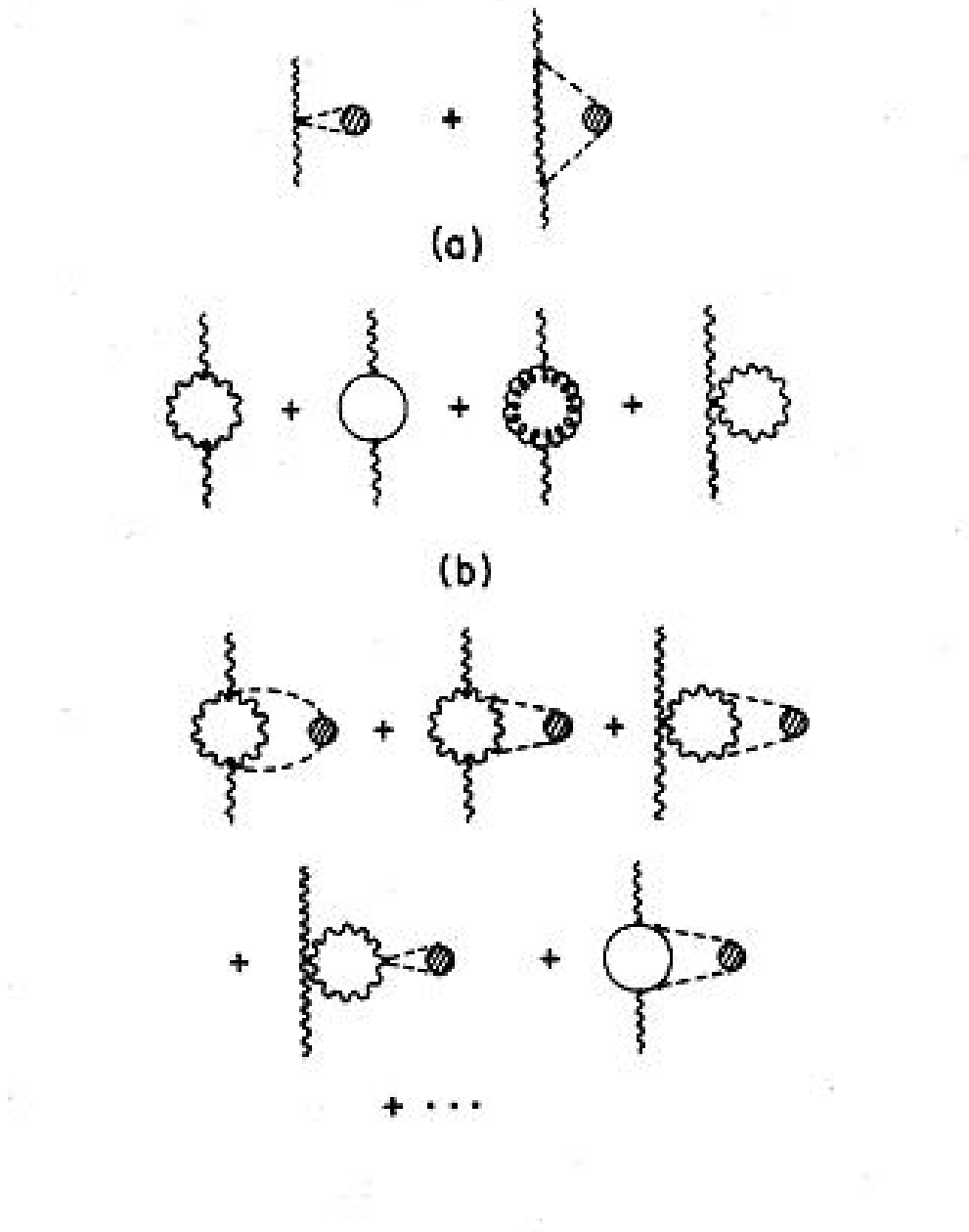}%
\caption{a) Calculation of the gluon self-energy in the
condensate. The first term shows the origin of the gluon mass term
in the mean-field approximation. The dashed line refers to a
condensate gluon of zero momentum. In the second part of a) we
show a contribution to the (irreducible) polarization tensor in
the single (condensate) loop approximation. b) Diagrams which
contribute to the polarization tensor in QCD. The wavy line is a
gluon, the solid line is a quark, and the third diagram represents
the gluon field. c) Some corrections to the diagrams of (b) due to
the presence of a gluon condensate.}
\end{figure}

It is also useful to subtract the quantity given in Eq. (4.1) from
the current and define \be
\hat{J}_a^\mu(k)=J_a^\mu(k)+m_G^2(g^{\mu\nu}-k^\mu
k^\nu/k^2)\mathcal{A}_\nu^a(k) \ee in momentum space. Then we have
\be \left[(k^2-m_G^2)(g^{\mu\nu}-k^\mu
k^\nu/k^2)+\frac{1}{\lambda}k^\mu
k^\nu\right]\,\mathcal{A}_\nu^a(k)=-\hat{J}_a^\mu(k). \ee

We now write a first-order propagator as \be
i\mathcal{D}_{ab}^{(1)\mu\nu}(k)=-i\,\left[\frac{g^{\mu\nu}-k^\mu
k^\nu/k^2}{k^2-m_G^2}+\lambda\frac{k^\mu
k^\nu}{(k^2+i\epsilon)^2}\right]\delta_{ab}, \ee and also write
\be
i\mathcal{D}(k^2)=i\mathcal{D}^{(1)}(k^2)+i\mathcal{D}^{(1)}(k^2)
\left[\frac{i\Pi_A(k^2)}{1-\frac{\Pi_A(k^2)}{k^2-m_G^2}}\right]i\mathcal{D}^{(1)}(k^2),
\ee where \be \mathcal{D}^{(1)}(k^2)=-[k^2-m_G^2]^{-1}, \ee and
\be \mathcal{D}(k^2)=-[k^2-m_G^2-\Pi_A(k^2)]^{-1}.\ee

Thus we see that, after absorbing the mass term in
$\mathcal{D}_{\mu\nu}^{(1)}$, the quantity $\Pi_A(k^2)$ is related
to the time-ordered product of the $\hat{J}_{T,\mu}^a(x)$: \be
[g^{\mu\nu}-k^\mu
k^\nu/k^2]\left[\frac{\Pi_A(k^2)}{1-\frac{\Pi_A(k^2)}{k^2-m_G^2}}\right]\delta_{ab}=i\,\int
d^4x\,e^{ik\cdot x}<vac\mid
T[\hat{J}_{T,a}^\mu(x)\hat{J}_{T,b}^\nu(0)]\mid vac>.\ee  Equation
(4.12) is a generalization of Eq. (2.18) and reflects the presence
of a condensate in the QCD vacuum which makes the gluon massive.
We also remark that, after gauge fixing, the theory with ghost
fields has a form of gauge invariance - the BRST gauge symmetry.
This symmetry allows one to derive the analog of the QCD Ward
identities in QCD - the Slavnov-Taylor identities. (We note that
the ghost condensate introduced in Ref. [26] is BRST invariant.)

\section{The gluon propagator and the Dielectric function in QCD}

One is tempted to write the analog of Eq. (2.1) in the case of
QCD. However, if there is a gluon condensate present, there is an
essential modification to be considered. We recall that we found
it useful to divide $A_\mu^a(x)$ into a condensate field,
$\mathbb{A}^a_\mu(x)$, and a fluctuating field,
$\mathcal{A}^a_\mu(x)$. (We made the assumption that the
condensate field is in the zero-momentum mode and therefore,
$\mathbb{A}_a^\mu(x)$ is independent of $x$.)

Thus, in coordinate space, we have \be
i\mathcal{D}_{\mu\nu}^{ab}(x,x')&=&<vac\mid
T[A_\mu^a(x)A_\nu^b(x')]\mid vac> \\ &=&<vac\mid
\mathbb{A}_\mu^a\mathbb{A}_\nu^b\mid vac>+<vac\mid T
[\mathcal{A}_\mu^a(x)\mathcal{A}_\nu^b(x')]\mid
vac>\\
&=&-\frac{g_{\mu\nu}}{4}\phi_0^2\frac{\delta_{ab}}{8}+<vac\mid T
[\mathcal{A}_\mu^a(x)\mathcal{A}_\nu^b(x')]\mid vac>. \ee

Our expression for the gluon propagator in momentum space is then
\be
i\mathcal{D}_{\mu\nu}^{ab}(k)=-\frac{g_{\mu\nu}}{4}\phi_0^2\frac{\delta_{ab}}{8}
(2\pi)^4\delta^{(4)}(k) -i\,\left[\frac{(g_{\mu\nu}-k_\mu
k_\nu/k^2)}{k^2[1-\Pi_1(k^2)]}+\lambda\frac{k_\mu
k_\nu}{(k^2+i\epsilon)^2}\right]. \ee We see the first
characteristic difference when we compare Eq. (5.4) with Eq.
(2.1), that is, the presence of a delta function. Whether such a
term is present depends on whether or not one has a ground-state
condensate in the zero-momentum mode.

We can define a QCD dielectric function: \be
\kappa(k^2)=[1-\Pi_1(k^2)].\ee As noted earlier, the Schwinger
mechanism refers to the a fact that, if $\kappa(k^2)$ has a pole
at $k^2=0$, the gluon has a dynamical mass and the pole at $k^2=0$
in $\mathcal{D}_{\mu\nu}^{ab}(k)$ disappears. In an earlier work
we found that $m_G^2=(9/32)g^2\phi_0^2$. As we will see in this
work \be
\kappa(k^2)=\left[1-\frac{m_G^2}{k^2}+\frac{4\eta^2}{k^2-m_G^2}\right]\ee
where $\eta^2=(3/32)g^2\phi_0^2$. (Thus $m_G^2=3\eta^2$.) The
result given in Eq. (5.6) follows from the calculation of the
diagrams in Fig. 1 subject to the constraints required in the
covariant formalism. The quantity $\kappa(k^2)$ defined in Eq.
(5.6) is shown in Fig. 2. It is interesting to note that if the
Schwinger mechanism is operative, that is, if there is a pole in
$\kappa(k^2)$ at $k^2=0$, one needs an additional singularity to
avoid having a zero in $\kappa(k^2)$. Such a zero would imply that
gluons could go on-mass-shell, a clearly unsatisfactory result. It
is gratifying that at the next level of approximation (one
condensate loop) one finds the necessary singular term that
maintains the relation $\kappa(k^2)\neq 0$.

\begin{figure}
\includegraphics[bb=120 50 400 400, angle=0, scale=1]{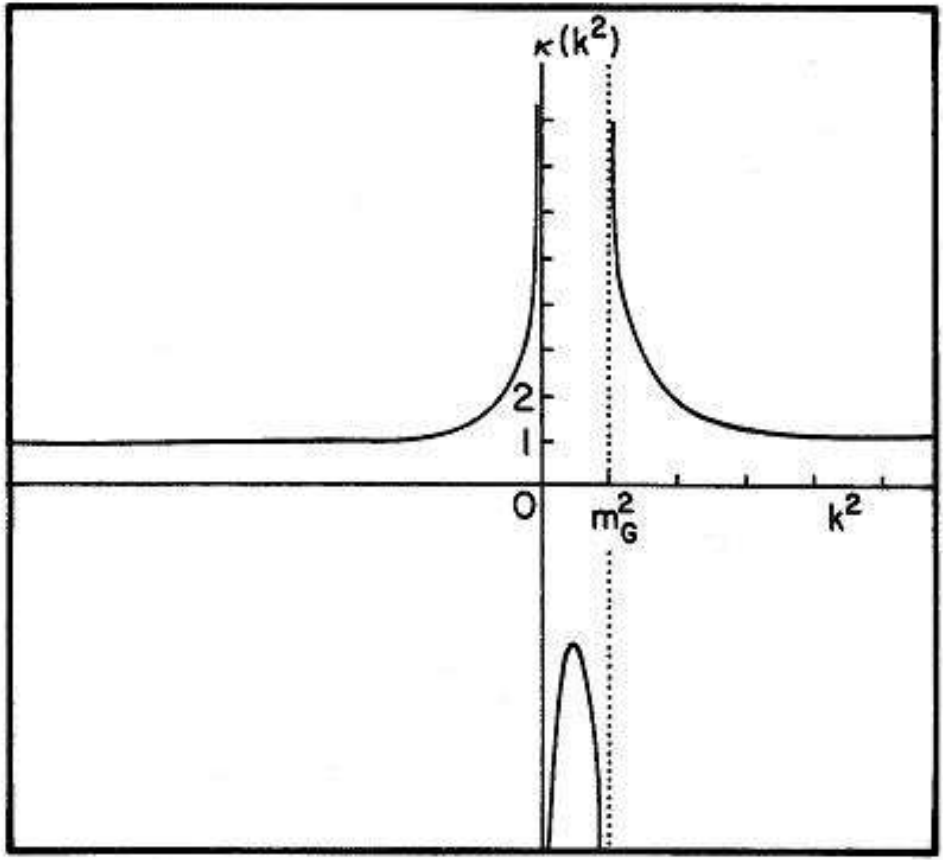}%
\caption{The dielectric constant
$\kappa(k^2)=[1-\frac{m_G^2}{k^2}+\frac{4\eta^2}{k^2-m_G^2}]$ is
shown. The singularity at $k^2=0$ reflects the operation of the
Schwinger mechanism [25]. (Here we see how the infrared
singularities of the theory lead to nonpropagation of gluons in
the QCD vacuum, a result which has long been conjectured to be
true.)}
\end{figure}

We do not have all the terms contributing to $\kappa(k^2)$. For
 example, there will be terms of order $g^2(\phi_0^2/k^2)$ or
 $g^2\ln(\phi_0^2/\mid k^2\mid)$ in the deep Euclidean region. The
 origin of such terms may be seen in Fig. 1, where we have shown
 how the presence of the condensate can lead to (power)
 corrections to the asymptotic behavior of the polarization tensor
 in the region $k^2\rightarrow -\infty$.

We also note that the only way to form a small parameter in this
model is to construct the ratio $[g^2\phi_0^2/(-k^2)]$ which is
small for large spacelike $k^2$. The nonperturbative analysis is
clearly not an expansion in a small parameter. That is
characteristic of nonperturbative approximations in general.
Usually it is difficult to find a completely satisfactory
organizational principle for a nonperturbative expansion. One that
is extensively used is a loop expansion. The zero-loop or
``tree-approximation" corresponds to the mean-field approximation.
This approximation is used extensively in field theory and
many-body physics. In our analysis the first term in Fig. 1a is
identified as the ``tree" or mean-field approximation. That
approximation is sufficient to generate the gluon mass via the
Schwinger mechanism. The second term in Fig. 1a may be thought of
as a (condensate) one-loop correction to the mean-field
approximation. It is, of course, interesting that we find
nonpropagation of gluons already in the one-loop approximation.
Increasing the number of condensate loops increases the number of
factors of $g^2\phi_0^2$ which appear in the numerator of the
terms which make up $\kappa(k^2)$. That is, at the tree level, we
obtain the $m_G^2/k^2$ term and at the one-loop level, we find the
term $-4\eta^2/(k^2-m_G^2)$.

We now wish to obtain the contribution to $\Pi^{\mu\nu}_{ab}(k)$
in the Landau gauge of the form \be (g^{\mu\nu}-k^\mu
k^\nu/k^2)\,\left[\frac{-4\eta^2k^2}{k^2-m_G^2}\right] \ee
displayed above. Various elements of our analysis are depicted in
Figs. 3-5 which are taken from Ref. [27].

Combining the above result with Eq. (4.3), the mean-field plus the
one (condensate) loop result for the polarization tensor is \be
\Pi_{\mu\nu}^{ab}(k) = \delta_{ab}(g_{\mu\nu}-k_\mu
k_\nu/k^2)\left[m_G^2 -\frac{-4\eta^2k^2}{k^2-m_G^2}\right].\ee

We had \be \delta_{ab}(g^{\mu\nu}-k^\mu
k^\nu/k^2)\left[\frac{\Pi(k^2)}{1-\frac{\Pi(k^2)}{k^2-m_G^2}}\right]
=i\int d^4x\,e^{ik\cdot x}<vac\mid
T[\hat{J}_{T,a}^\mu(x)\hat{J}_{T,b}^\nu(0)]\mid vac>. \ee

From our definition of $J_\nu^a(x)$, we find \be
J_\nu^a(x)=gf^{abc}[A_b^\mu(x)\partial_\nu
A_\mu^c(x)-2A_b^\mu(x)\partial_\mu
A_\nu^c(x)-A_\nu^c(x)\partial^\mu A_\mu^b(x)]\\\nonumber
+g^2f^{abc}f^{a'b'c}[A_\mu^b(x)A_{b'}^\mu(x)A_\nu^{a'}(x)]. \ee We
insert \be A_\mu^a(x) =\mathbb{A}_\mu^a(x)+\mathcal{A}^a_\mu(x)
\ee into the last expression to obtain \be
J_\nu^a(x)=gf_{abc}[\mathbb{A}_b^\mu(x)\partial_\nu
\mathcal{A}_\mu^c(x)-2\mathbb{A}_b^\mu(x)\partial_\mu
\mathcal{A}_\nu^c(x)
-\mathbb{A}_\nu^c(x)\partial^\mu\mathcal{A}_\mu^b(x)]
\\\nonumber
+gf_{abc}[\mathcal{A}_b^\mu(x)\partial_\nu\mathcal{A}_\mu^c(x)-
2\mathcal{A}_b^\mu(x)\partial_\mu\mathcal{A}_\nu^c(x)-
\mathcal{A}_\nu^c(x)\partial^\mu\mathcal{A}_\mu^b(x)] \\\nonumber
+g^2f^{abc}f^{a'b'c}[\mathbb{A}_\mu^b(x)+\mathcal{A}_\mu^b(x)]
[\mathbb{A}^\mu_{b'}(x)+\mathcal{A}^\mu_{b'}(x)]
[\mathbb{A}_\nu^{a'}(x)+\mathcal{A}_\nu^{a'}(x)].\ee

Since we are here working to order $(g^2\phi_0^2)$, we will drop
the last term of Eq. (5.12) at this point. (However, we note that
it is responsible for the term proportional to $m_G^2$ in Eq.
(5.8).) Thus, we may use the approximation \be \bar{J}_\nu^a(x)
\cong gf^{abc}[\mathbb{A}_b^\mu \partial_\nu
\mathcal{A}_\mu^c(x)-2\mathbb{A}^\mu_b\partial_\mu\mathcal{A}_\nu^c(x)
+\mathbb{A}_\nu^b\partial^\mu\mathcal{A}_\mu^c(x)]\ee for the
calculation to be made here. (Note that, in the last term, we have
interchanged $b$ and $c$ and changed the sign of that term.) We
maintain the constraint \be <\psi_m\mid
\partial_\mu\bar{J}_a^\mu(x)\mid\psi_n>=0, \ee and implement that
constraint by using only the conserved current,
$\hat{J}_{T,\mu}^a(x)$, in our calculation. We can define the
projection operator \be \mathcal{P}_{\mu\nu}^T
=g_{\mu\nu}-\partial_\mu\frac{1}{\Box}\partial_\nu, \ee which in
momentum space has the form \be \mathcal{P}_{\mu\nu}^T
=[g_{\mu\nu}-k_\mu k_\nu/k^2]. \ee We have \be
\hat{J}_{T,\nu}^a(x)\equiv \mathcal{P}_{\nu\mu}^TJ_a^\mu(x). \ee

\begin{figure}
\includegraphics[bb=25 25 400 420, angle=0, scale=1]{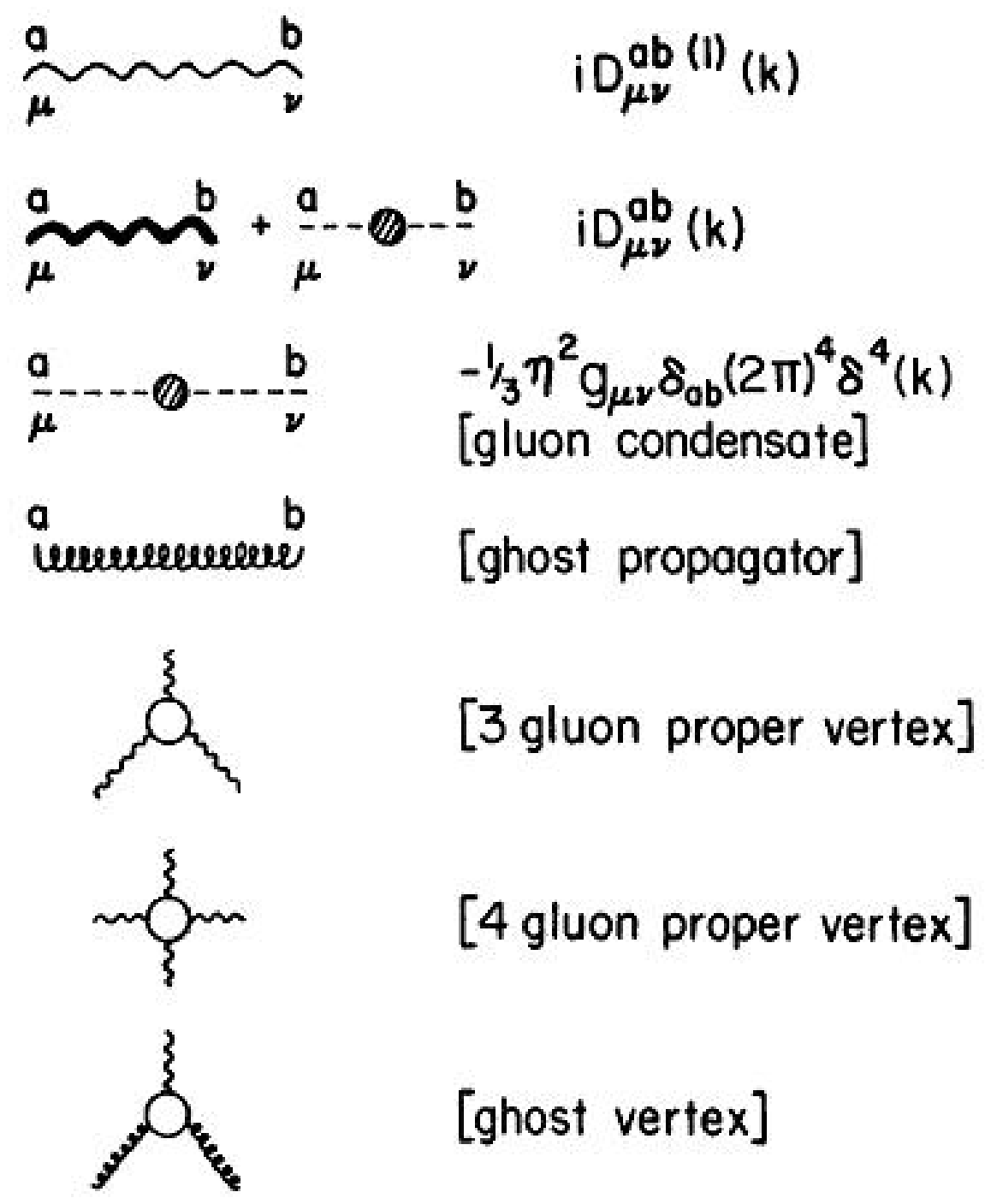}%
\caption{Diagrammatic representation of various elements of our
model. Note that the dashed line contributes for $k_\mu=0$. (See
Ref. [27] for further details.)}
\end{figure}

\begin{figure}
\includegraphics[bb=80 270 500 420, angle=0, scale=1]{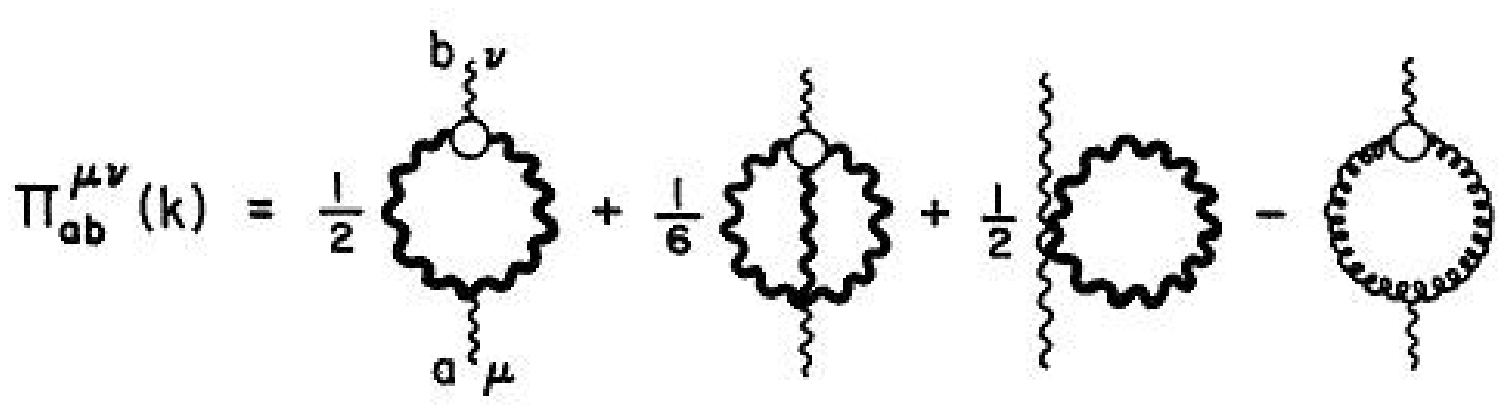}%
\caption{Diagrammatic representation of the equations determining
the vacuum polarization tensor. (See Ref. [27].)}
\end{figure}

\begin{figure}
\includegraphics[bb=30 225 500 460, angle=0, scale=1]{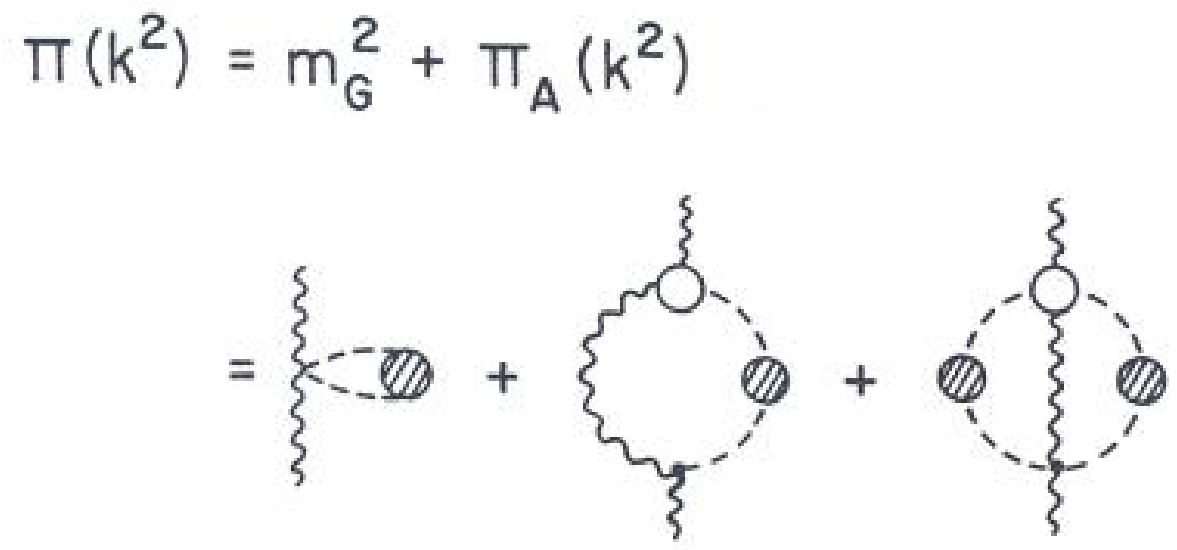}%
\caption{Diagrammatic representation of various elements of the
polarization tensor used in this work. The first diagram on the
right-hand side is responsible for the gluon mass term. [The
vertex functions in the second two terms are to be expressed in
terms of the full gluon propagator,
$i\mathcal{D}_{\mu\nu}^{ab}(k)$.]}
\end{figure}

Thus, using Eq. (5.13), we find \be \hat{J}_{T,\nu}^a(x) \simeq
gf^{abc}[\mathbb{A}_b^\rho P_{\nu\mu}^T \partial^\mu
\mathcal{A}_\rho^c(x)-2\mathbb{A}_b^\rho
P_{\nu\mu}^T\partial_\rho\mathcal{A}_c^\mu(x)
\\\nonumber
+P^T_{\nu\mu}{\mathbb{A}_b^\mu\partial^\rho\mathcal{A}_\rho^c(x)}].
\ee The first term in Eq. (5.18) is equal to zero. the last term
in Eq. (5.18) can be dropped because of the constraint \be
<\psi_n\mid \partial_\mu\mathcal{A}_c^\mu(x)\mid\psi_m>=0.\ee To
the order considered, we have  \be \nonumber i\,&<& vac\mid
T[\hat{J}_{T,a}^\nu(x)\hat{J}_{T,a'}^{\nu'}(x')]\mid
vac>\qquad \qquad \qquad \qquad\qquad\qquad\qquad\qquad\qquad\qquad\,\,\,\,\,\,\,\,\,\\
&=& i\,g^2f^{abc}f^{a'b'c}<vac\mid
T\{-2\mathbb{A}_\rho^bP_T^{\nu\mu}\,\partial^\rho\mathcal{A}_\mu^c(x)\}
\times\{-2\mathbb{A}_{\rho'}^{b'}P_T^{\nu'\mu'}\partial^{'\rho'}\mathcal{A}_{\mu'}^{c'}(x')\}\mid
vac>\\
&=&i\,g^2f^{abc}f^{a'b'c}<vac\mid\mathbb{A}_\rho^b\,\mathbb{A}_{\rho
'}^{b'}\mid vac>\\\nonumber &&
\times\{4P_T^{\nu\mu}\,P_T^{\nu'\mu'}\}<vac\mid T
\{\partial^\rho\mathcal{A}_\mu^c(x)\,\partial '^{\rho
'}\mathcal{A}_{\mu'}^{c'}(x')\}\mid vac>\\
&=&i\,g^2f^{abc}f^{a'b'c}\left[-\phi_0^2\frac{\delta_{bb'}}{8}\frac{g_{\rho'\rho}}{4}\right]
\{4P_T^{\nu\mu}P_T^{\nu'\mu'}\}\,\partial^\rho\partial '^{\rho
'}\,[i\,\mathcal{D}_{\mu\mu'}^{cc'}(x,x')]. \ee

We write \be
\mathcal{D}_{\mu'\nu'}^{cc'}(x,x')=\delta_{cc'}\mathcal{D}_{\mu\nu'}(x,x'),\ee
and find \be  i\,< vac\mid T
[\hat{J}_{T,a}^\nu(x)\hat{J}_{T,a'}^{\nu'}(x')]\mid
vac>=\frac{3}{32}g^2\phi_0^2\,\delta_{aa'}\,[\,4P_T^{\nu\mu}P_T^{\nu'\mu'}\partial_\rho\partial^{'\rho}\,\mathcal{D}_{\mu\mu'}(x,x')].
\ee

Now introduce \be
\mathcal{D}_{\mu\mu'}(k)=-\left[\frac{P^T_{\mu\mu'}}{k^2-m_G^2-\Pi_A(k^2)}+\frac{\lambda
k_\mu k_{\mu'}}{(k^2+i\epsilon)^2}\right], \ee and note that
$\partial^\rho\partial^{'}_\rho\rightarrow[ik^\rho][-ik_\rho]=k^2$.

Further, $P_T^{\nu\mu}P_T^{\nu'\mu'}P_{\mu\mu'}^T=P_T^{\nu\nu'}$,
and therefore, \be  \delta_{aa'}(g^{\nu\nu'}-k^\nu
k^{\nu'}/k^2)\left[\frac{\Pi_A(k^2)}{1-\frac{\Pi_A(k^2)}{k^2-m_G^2}}\right]=(g^{\nu\nu'}-k^\nu
k^{\nu'}/k^2)\left[\frac{-4\eta^2k^2}{k^2-m_G^2-\Pi(k^2)}\right]\delta_{aa'},
\ee where $\eta^2=(3/32)g^2\phi_0^2$.

Thus \be
\frac{\Pi_A(k^2)}{[1-\frac{\Pi_A(k^2)}{k^2-m_G^2}]}=-\left[\frac{-4\eta^2k^2}{k^2-m_G^2-\Pi_A(k^2)}\right],\ee
which has the solution, \be
\Pi_A(k^2)=\frac{-4\eta^2k^2}{k^2-m_G^2}.\ee Recall that \be
\Pi(k^2) = m_G^2 +\Pi_A(k^2), \ee which then yields Eq. (5.7).

One may ask how our result is related to the result of a
diagrammatic analysis. It may be seen that the result given here
is obtained if one calculates in the Landau gauge and adds a ghost
condensate to maintain the transverse structure for
$\Pi_{\mu\nu}(k)$. Indeed, Lavelle and Schaden [26] have used a
ghost condensate to enforce the transverse nature of the
nonperturbative part of $\Pi_{\mu\nu}(k)$. (Recall Eq. (2.26).)
Their calculation is made in the deep-Euclidean region
$(k^2\rightarrow-\infty)$. Therefore, we can compare our result
with theirs, in the case a condensate $<:A^2:>$ is present, by
taking $k^2\rightarrow-\infty$ in our result. From Eqs. (5.28) and
(5.29) we have \be \Pi(k^2)&\longrightarrow&
m_G^2-\frac{4}{3}m_G^2\qquad
k^2\rightarrow-\infty,\\&=&-\frac{1}{3}m_G^2,\\&=&-\frac{3}{32}g^2(\tilde{\mu}^2)\phi_0^2,\ee
which agrees with the result of Ref. [26], when that result is
evaluated in the Landau gauge. (It is interesting to see how the
sign of $\Pi(k^2)$ changes as one passes from $k^2=0$ to
$k^2=-\infty$.)

\section{QCD lattice calculations and phenomenological forms for the Euclidean-space gluon propagator}

The form we obtained for the propagator was \be
D^{\mu\nu}(k)=\left(g^{\mu\nu}-\frac{k^\mu k^\nu}{k^2}\right)D(k).
\ee We now write \be
D(k)=\frac{Z_1}{k^2-m^2+\frac{4}{3}\frac{k^2m^2}{k^2-m^2}}.\ee
Here $Z_1$ is a normalization parameter which we put equal to 3.82
so that we may obtain a continuous representation as we pass from
Minkowski to Euclidean space.  In Fig. 6 we show $D(k)$ with
$m^2=0.25$ GeV$^2$. (We remark that $D(k)=0$ when $k^2=m^2$,
$D(k)=-Z_1/m^2$ at $k^2=0$, and $D(k)\rightarrow Z_1/k^2$ for
large $k^2$.) If we choce $Z_1=15.28m^2=3.82$ our result for the
propagator will be continuous at $k^2=0$ when we consider both the
Euclidean-space and Minkowski-space propagators.

Results for the gluon propagator obtained in a lattice simulation
of QCD are given in Ref. [28]. In that work the authors also
record several phenomenological forms. We reproduce these forms in
the Appendix for ease of reference. Of these various forms we will
make use of model A of Ref. [28] which has the form \be
D^L(k^2)=Z\left[\frac{AM^{2\alpha}}{(k^2+M^2)^{1+\alpha}}+\frac{1}{k^2+M^2}L(k^2,M)\right],\ee
with \be L(k^2,M)\equiv
\left[\frac{1}{2}\ln(k^2+M^2)(k^{-2}+M^{-2})\right]^{-d_D},\ee and
$d_D=13/22$. The parameters used in Ref. [28] to provide a very
good fit to the QCD lattice data are \be Z=2.01^{+4}_{-5},\ee \be
A=9.84^{+10}_{-86},\ee \be M=0.54^{+5}_{-5},\ee and \be
\alpha=2.17^{+4}_{-19}.\ee Note that $M$ in GeV units is $1.018$
GeV. Rather than work with the lattice data we will use Eqs.
(6.3)-(6.8) when we compare our results with the lattice data. In
Fig. 7 we show $k^2D^L(k)$ of Eq. (6.3) and in Fig. 8 we show
$D^L(k)$. These functions are represented by the solid lines in
Figs. 7 and 8. Note that Eq. (6.2) may be written in Euclidean
space as \be
D_E(k)=-\frac{Z_1}{k^2_E+m^2-\frac{4}{3}\frac{k^2_Em^2}{k^2_E+m^2}}.\ee
This form is useful for $k_E^2<1$ GeV$^2$ and we therefore
consider various phenomenological forms which may be used to
extend Eq. (6.9) so that we may attempt to fit the lattice result
over a broader momentum range. To that end, we make use of Ref.
[29]. The authors of that work define the Landau gauge gluon
propagator as \be
<A^a_\mu(k)A^a_\nu(k')>=V\delta(k+k')\delta^{ab}\left(\delta_{\mu\nu}-\frac{k_\mu
k_\nu}{k^2}\right)\frac{Z(k^2)}{k^2}, \ee with \be
Z(k^2)=\omega\left(\frac{k^2}{\Lambda^2_{QCD}+k^2}\right)^{2\kappa}(\alpha(k^2))^{-\gamma},\ee
and $\gamma=-13/22$. (We do not ascribe any particular
significance to Eq. (6.11). We use Eq. (6.11) as a
phenomenological form which could be replaced by a form which
provides a better fit to the data within the context of our model
at some future time. We believe Eq. (6.11) is useful, since it is
a simple matter to remove the first term of that equation and
introduce a propagator that has the small $k^2$ behavior of our
model.)

The authors of Ref. [29] introduce two choices for $\alpha(k^2)$
of Eq. (6.11). We use their form for $\alpha_2(k^2)$: \be
\alpha_2(k^2)=\frac{\alpha(0)}{\ln\left[e+a_1\left(\frac{k^2}{\Lambda^2_{QCD}}\right)^{a_2}\right]}.\ee
In their analysis they put $\kappa=0.5314$, $\Lambda_{QCD}=354$
MeV, $\alpha(0)=2.74$, $a_1=0.0065$ and $a_2=2.40$. (Here, we have
not recorded the uncertainties in these values which are given in
Table 2 of Ref. [29].) As we proceed, we will change these values
somewhat. As a first step we remove the first factor in Eq. (6.11)
and write \be Z(k^2)=Z_2(\alpha_2(k^2))^{-\gamma}.\ee We now use
$a_1=0.0080$ and $a_2=2.10$ rather than the values given above. In
Fig. 9 we show $(\alpha_2(k))^{13/22}$ as a function of $k$, using
our modified values of $a_1$ and $a_2$.

We now define \be
D_E(k_E)=-\frac{Z_2(\alpha(k^2))^{-\gamma}}{k_E^2+m^2-\frac{4}{3}\frac{k_E^2m^2}{k_E^2+m^2}}.\ee
The function $-k^2D_E(k_E)$ is shown in Fig. 7 as a dotted line.
In this calculation we have put $Z_2=2.11$. We find a good
representation of the lattice result for $k_E<2$ GeV,

In Fig. 8 we compare $D_E(k_E)$ with the result of the lattice
calculation which is represented by the solid line. In Fig. 10 we
combine our results in Minkowski and Euclidean space and show the
values of $k^2D(k^2)$ for both positive and negative $k^2$ values.
For positive $k^2$ we use $D(k)$ of Eq. (6.2) and for negative
values of $k^2$ we use $D_E(k_E^2)$ of Eq. (6.14). Equality of
these functions at $k^2=0$ implies $Z_1=Z_2(\alpha(0))^{13/22}$,
or $Z_1=1.81Z_2$. (In our work we have used $Z_1=3.82$ and
$Z_2=2.11$. See Eqs. (6.2) and (6.14).) In Fig. 11 we show
$D(k^2)$ rather than $k^2D(k^2)$, which was shown in Fig. 10.

\begin{figure}
\includegraphics[bb=40 25 200 200, angle=0, scale=1]{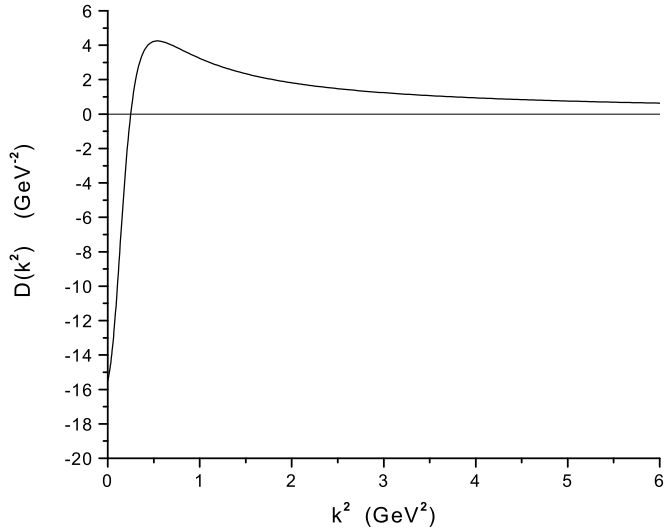}%
\caption{The function $D(k^2)$ of Eq. (6.2) is shown in Minkowski
space. The value for large $k^2$ is given by $Z_1/k^2$ with
$Z_1=3.87$. Here $m=0.50$ GeV.}
\end{figure}

\begin{figure}
\includegraphics[bb=40 25 240 240, angle=0, scale=1]{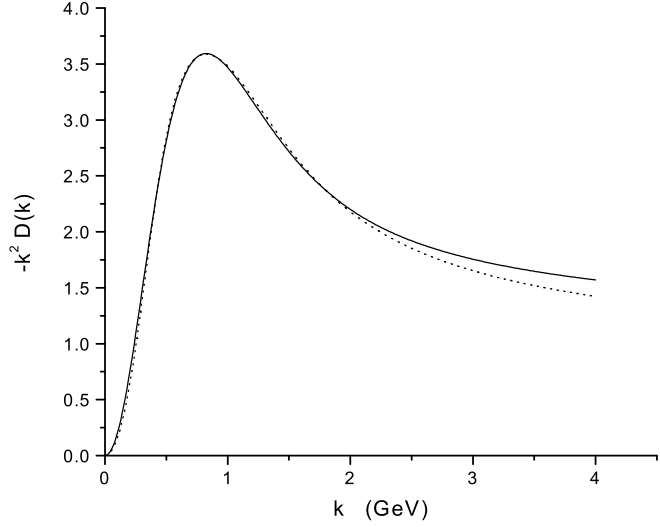}%
\caption{The function $-k_E^2D_E(k)$ is shown. The solid line
represents the QCD lattice data, while the dotted line represents
$-k^2_ED_E(k)$ in the case that $D_E(k)$ is given in Eq. (6.14).}
\end{figure}

\begin{figure}
\includegraphics[bb=40 25 240 240, angle=0, scale=1]{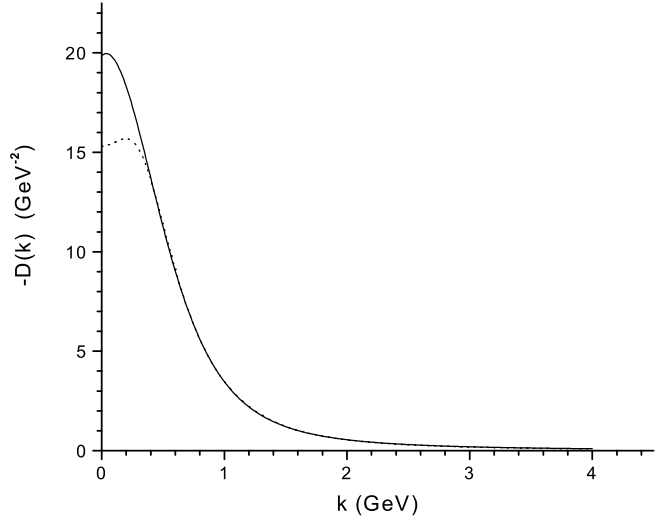}%
\caption{The function $-D_E(k)$ is shown. The solid line
represents the QCD lattice data, while the dotted line represents
$-D_E(k)$ of Eq. (6.14). [See Fig. 7.]}
\end{figure}

\begin{figure}
\includegraphics[bb=40 25 240 240, angle=0, scale=1]{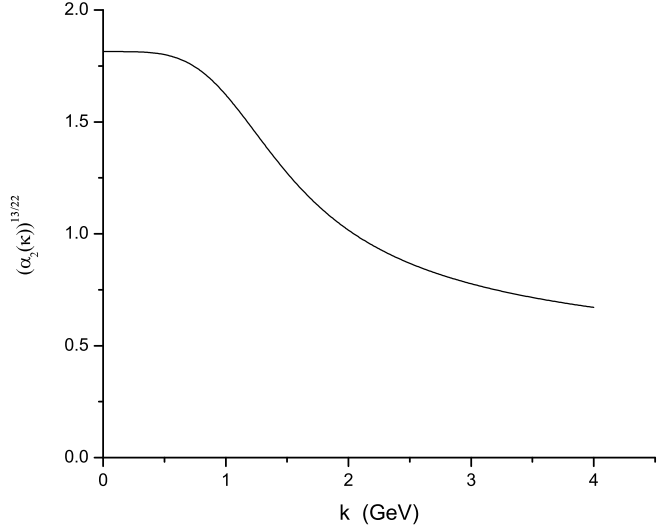}%
\caption{The function $(\alpha_2(k))^{13/22}$ is shown. [See Eq.
(6.12).] Note that $(\alpha_2(0))^{13/22}=1.81$.}
\end{figure}

\begin{figure}
\includegraphics[bb=40 25 240 240, angle=0, scale=1]{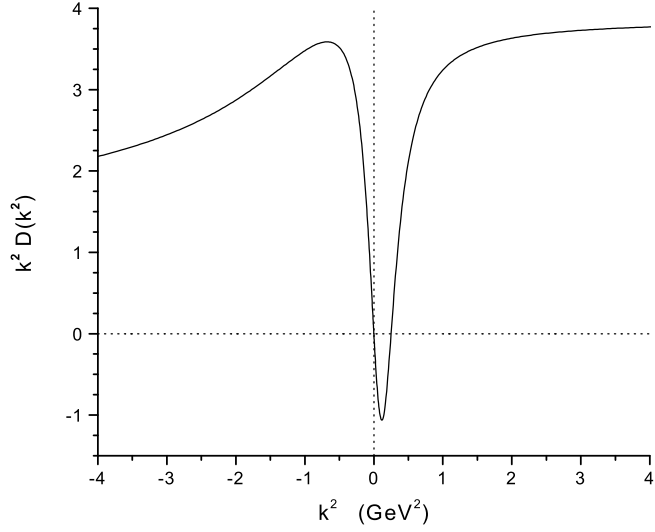}%
\caption{For $k^2>0$ the solid line represents $k^2D(k^2)$ with
$D(k^2)$ given by Eq. (6.2). Here, $Z_1=3.82$. For $k^2<0$ we show
$k^2D_E(k^2)$, where $D_E(k^2_E)$ is given by Eq. (6.14) with
$Z_2=2.11$.}
\end{figure}

\begin{figure}
\includegraphics[bb=40 25 240 240, angle=0, scale=1]{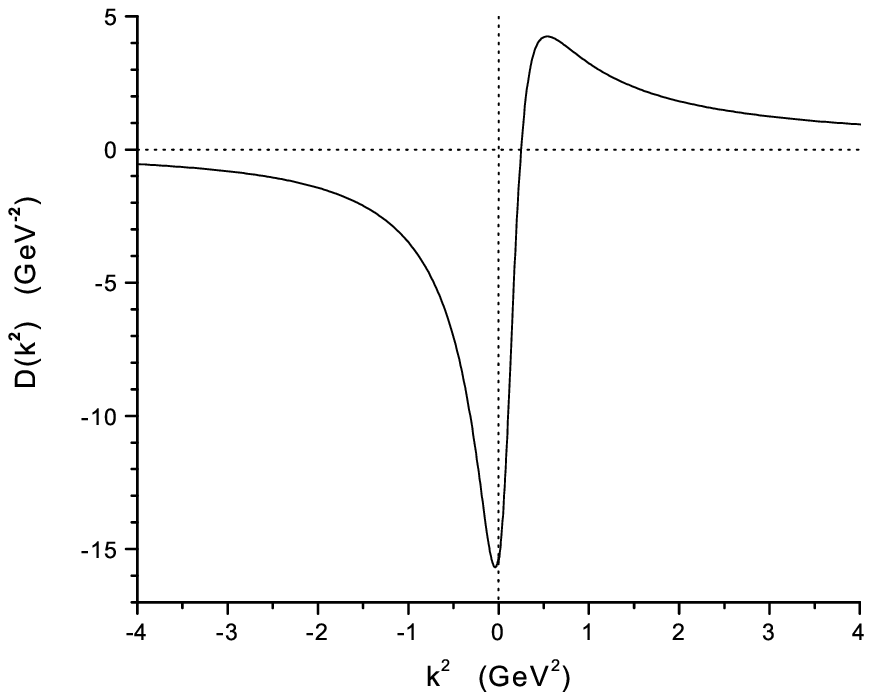}%
\caption{Same as Fig. 10 except that $D(k^2)$ is shown.}
\end{figure}

\section{discussion}

In this work we have developed nonperturbative approximations for
the description of the gluon condensate and have calculated the
form of the gluon propagator. The approximation used may be
thought of as a condensate-loop expansion. Since the condensate is
assumed to be in the zero-momentum mode, the loop expansion does
not require loop integrals, but leads to algebraic relations. Our
results are obtained in the Landau gauge. (Note that ghosts are
introduced to maintain the transverse character of
$\Pi_{\mu\nu}(k)$ in Ref. [26].) We are able to make some contact
with lattice calculations of the gluon propagator, which are made
in the Landau gauge. We find that our value for the dynamical
gluon mass, $m_G\simeq$ 600 MeV, is in accord with the results of
recent lattice calculations. We have also seen that our results
agree with those of Lavelle and Schaden, if one evaluates our
propagators in the deep-Euclidean region $(k^2\rightarrow-\infty)$
[26].

In our work, confinement of quarks and gluons and chiral symmetry
breaking are related to a single condensate order parameter
$(g^2\phi_0^2)$. This result is consistent with the fact that in
lattice simulations of QCD, deconfinement and chiral symmetry
restoration take place at the same temperature. In this
connection, we note that there is no threshold value of
$(g^2\phi_0^2)$. For any finite value of this parameter, we find
chiral symmetry breaking and nonpropagation of quarks [30] and
gluons.

In this work we have provided a representation of the gluon
propagator in both Euclidean and Minkowski space. The
Minkowski-space propagator has only complex poles and that implies
that the gluon is a nonpropagating mode in the QCD vacuum. Our
analysis takes into account the important condensate $<A_a^\mu
A_\mu^a>$ which is responsible for mass generation for the gluon.
Our work has some relation to that of Cornwall [31] who obtained a
gluon mass of $500\pm 200$ MeV in his analysis. Cornwall also
suggested that ``quark confinement arises from a vertex condensate
supported by a mass gap."

In recent work, Gracey obtained a pole mass of the gluon of
$2.13\Lambda_{\overline{MS}}$ in a two-loop renormalization scheme
[32]. If we put $\Lambda_{\overline{MS}}=250$ MeV, the mass
obtained at two-loop order in Ref. [32] is $532$ MeV, which is
close to the value of $500$ MeV used in the present work. (We
remark that in Ref. [22] we obtained a gluon mass of $530$ MeV, if
we made use of Eq. (3.18) of that reference, which includes the
effect of including various exchange terms in our analysis of the
relevant matrix elements.)

\appendix
  \renewcommand{\theequation}{A\arabic{equation}}
  \setcounter{equation}{0}  
  \section{}  

For ease of reference we record various semi-phenomenological
forms which are meant to represent the Euclidean-space gluon
propagator.

Gribov [33]: \be D^L(k^2)=\frac{Zk^2}{k^4+M^4}L(k^2,M).\ee Stingl
[34]: \be D^L(k^2)=\frac{Zk^2}{k^4+2A^2k^2+M^4}L(k^2,M).\ee
Marenzoni et al. [35]: \be
D^L(k^2)=\frac{Z}{(k^2)^{1+\alpha}+M^2}.\ee Cornwall I [31]: \be
D^L(k^2)=Z\left[[k^2+M^2(k^2)]\ln\left(\frac{k^2+4M^2(k^2)}{\Lambda^2}\right)\right]^{-1},
\ee where \be M(k^2)=M
\left[\frac{\ln\left(\frac{k^2+4M^2}{\Lambda^2}\right)}{\ln\left(\frac{4M^2}{\Lambda^2}\right)}\right]^{-6/11}.\ee
Cornwall II [36]: \be
D^L(k^2)=Z\left[[k^2+M^2]\ln\left(\frac{k^2+4M^2}{\Lambda^2}\right)\right]^{-1}.
\ee Cornwall III [36]: \be
D^L(k^2)=\frac{Z}{k^2+Ak^2\ln\left(\frac{k^2}{M^2}\right)+M^2}.\ee
Model A [28]: \be D^L(k^2)=
Z\left[\frac{AM^{2\alpha}}{(k^2+M^2)^{1+\alpha}}+\frac{1}{k^2+M^2}L(k^2,M)\right].\ee
The parameters for model A are given in Eqs. (6.5)-(6.7).\\Model B
[28]: \be D^L(k^2)=
Z\left[\frac{AM^{2\alpha}}{(k^2)^{1+\alpha}+(M^2)^{1+\alpha}}+\frac{1}{k^2+M^2}L(k^2,M)\right].\ee
Model C [28]: \be D^L(k^2)=
Z\left[\frac{A}{M^2}e^{-(k^2/M^2)^\alpha}+\frac{1}{k^2+M^2}L(k^2,M)\right].\ee


\vspace{1.5cm}
\begin{thebibliography}{99}

    \bibitem{B1}Ph. Boucaud, A. Le Yaouanc, J. P. Leroy, J.
    Micheli, O. P$\grave{\mbox{e}}$ne and J. Rodriguez-Quintero, Phys. Rev. D \textbf{63},
    114003 (2001).
    \bibitem{B2}Ph. Boucaud, A. Le Yaouanc, J. P. Leroy, J.
    Micheli, O. P$\grave{\mbox{e}}$ne and J. Rodriguez-Quintero, Phys. Lett. B \textbf{493},
    315 (2000).
    \bibitem{B3}E. R. Arriola, P. O. Bowman, and W. Broniowski,
    hep-ph/0408309, v3.
    \bibitem{B4}Ph. Boucaud, J. P. Leroy, A Le Yaouanc, J. Micheli, O. P$\grave{\mbox{e}}$ne, F. DeSoto, A. Donini, H. Moutarde and
J. Rodriguez-Quintero, Phys. Rev. D \textbf{66}, 034504 (2002).
    \bibitem{B5} B. M. Gripaios, Phys. Lett. B \textbf{558}, 250
    (2003).
    \bibitem{B6}K.-I. Kondo, Phys. Lett. B \textbf{572},
    210(2003).
    \bibitem{B7}K.-I. Kondo, Phys. Lett. B \textbf{514},
    335(2001).
    \bibitem{B8}A. A. Slavov, hep-th/0407194.
    \bibitem{B9} L. Stodolsky, Pierre van Baal and V. I. Zakharov,
    Phys. Lett. B \textbf{552}, 214(2002).
    \bibitem{B10}F. V. Gubarev, L. Stodolsky, and V. I. Zakharov, Phys. Rev.
        Lett. \textbf{86}, 2220 (2001).
    \bibitem{B11}F. V. Gubarev, V. I. Zakharov, Phys. Lett. B \textbf{501}, 28(2001).
    \bibitem{B12}J. A. Gracey, Phys. Lett. B \textbf{552}, 101 (2003). See also D. Dudal, H. Verschelde and S. P. Sorella, Phys.
    Lett. B \textbf{555}, 126 (2003).
    \bibitem{B13}D. Dudal, H. Verschelde, R. E. Browne and J. A. Gracey, Phys.
    Lett. B \textbf{562}, 87(2003).
    \bibitem{B14}D. Dudal, H. Verschelde, V. E. R. Lemes, M. S.
    Sarandy, R. F. Sobreiro, S. P. Sorella, M. Picariello, A.
    Vicini and J. A. Gracey, hep-th/0308153.
    \bibitem{B15} H. Verschelde, K. Knecht, K. Van Acoleyen and V. Vanderkelen, Phys. Lett. B \textbf{516}, 307(2001).
   \bibitem{B16}D. Dudal, H. Verschelde, V. E. R. Lemes, M. S.
    Sarandy, R. F. Sobreiro, S. P. Sorella and J. A. Gracey, Phys.
    Lett. B \textbf{574}, 325(2003).
   \bibitem{B17}D. Dudal, H. Verschelde, J. A. Gracey, V. E. R. Lemes, M. S.
    Sarandy, R. F. Sobreiro and S. P. Sorella, JHEP \textbf{0401},
    044(2004); hep-th/0311194, v3.
    \bibitem{B18}R. E. Browne and J. A. Gracey, JHEP \textbf{0311},
    029(2003); hep-th/0306200.
   \bibitem{B19}D. Dudal, J. A. Gracey, V. E. R. Lemes, R. F. Sobreiro, S. P.
   Sorella and H. Verschelde, hep-th/0409254.
    \bibitem{B20} B. L. Ioffe, Phys. Atom. Nucl. \textbf{66}, 30 (2003)
    \bibitem{B21} B. L. Ioffe and K. N. Zyablyuk, Eur. Phys. J. C \textbf{27},
229 (2003)
   \bibitem{B22}L. S. Celenza and C. M. Shakin, Phys. Rev. D
   \textbf{34}, 1591(1986)
   \bibitem{B23}M. A. Shifman, A. I. Vainstem and V. I.
      Zakharov, Nucl. Phys. B \textbf{147}, 385 (1979); B
      \textbf{147}, 448(1979); B \textbf{147}, 519(1979).
   \bibitem{B24}L. S. Celenza, Chueng-Ryong Ji and C. M. Shakin, Phys. Rev. D
   \textbf{36}, 895(1987)
 \bibitem{B25}J. Schwinger, Phys. Rev. \textbf{125}, 397(1962).
  \bibitem{B26}M. J. Lavelle and M. Schaden, Phys. Lett.
  \textbf{208}, 297(1988).
   \bibitem{B27}E. J. Eichten and G. L. Feinberg, Phys. Rev. D
   \textbf{10}, 3254(1974).
  \bibitem{B28}D. B. Leinweber, J. I. Skullerud, A. G. Williams
   and C. Parrinello, Phys. Rev. D \textbf{60}, 094507(1999).
   \bibitem{B29}O. Oliveira and P. J. Silva, hep-lat/0410048.
\bibitem{B30} Xiangdong Li and C. M. Shakin, Phys. Rev. D \textbf{70},
114011 (2004)
   \bibitem{B31}J. Cornwall, Phys. Rev. D \textbf{26}, 1453(1982).
   \bibitem{B32}J. A. Gracey, hep-ph/0411169.
   \bibitem{B33}V. N. Gribov, Nucl. Phys. B \textbf{139},
   19(1978).
   \bibitem{B34}M. Stingl, Phys. Rev. D \textbf{34}, 3863(1986);
   \textbf{36}, 651(1987).
   \bibitem{B35}P. Merenzoni, G. Martinelli and N. Stella, Nucl.
   Phys. B \textbf{455}, 339(1995); P. Marenzoni, G. Martinelli, N.
   Stella and M. Testa, Phys. Lett. B \textbf{318}, 511(1993).

   \bibitem{B36}J. Cornwall (private communication to the authors
   of Ref.[28]).


\end{thebibliography}

\end{document}